\documentclass[conference]{IEEEtran}
\IEEEoverridecommandlockouts
\usepackage{cite}
\usepackage{amsmath,amssymb,amsfonts}
\usepackage{algorithmic}
\usepackage{graphicx}
\usepackage{textcomp}
\usepackage{xcolor}
\usepackage{tikz}
\usetikzlibrary{arrows, arrows.meta}
\usepackage{setspace}
\usepackage{booktabs}
\usepackage{url}
\usepackage{flushend}
\usepackage{fancyhdr}

\def\BibTeX{{\rm B\kern-.05em{\sc i\kern-.025em b}\kern-.08em
    T\kern-.1667em\lower.7ex\hbox{E}\kern-.125emX}}

\fancypagestyle{firstpage}{
  \setlength{\headheight}{3.5\normalbaselineskip}
  \setlength{\headsep}{1.5\normalbaselineskip}

  \fancyhead[C]{%
    \footnotesize%
    Extended version of the paper to be presented at IEEE MIPR 2025, San Jose, CA, USA, August 2025 \\
    \vspace{0.8\normalbaselineskip}%
  }
  \fancyfoot{}
}

\begin{document}

\title{Rate-Accuracy Bounds in Visual Coding for Machines}

\author{\IEEEauthorblockN{Ivan V. Baji\'{c}\thanks{This work was supported by the NSERC grant RGPIN-2021-02485.}}
\IEEEauthorblockA{\textit{School of Engineering Science} \\
\textit{Simon Fraser University}\\
Burnaby, BC, Canada 
}
}

\maketitle

\begin{abstract}
Increasingly, visual signals such as images, videos and point clouds are being captured solely for the purpose of automated analysis by computer vision models. Applications include traffic monitoring, robotics, autonomous driving, smart home, and many others. This trend has led to the need to develop compression strategies for these signals for the purpose of analysis rather than reconstruction, an area often referred to as ``coding for machines.'' By drawing parallels with lossy coding of a discrete memoryless source, in this paper we derive rate-accuracy bounds on several popular problems in visual coding for machines, and compare these with state-of-the-art results from the literature. The comparison shows that the current results are at least an order of magnitude -- and in some cases two or three orders of magnitude -- away from the theoretical bounds in terms of the bitrate needed to achieve a certain level of accuracy. This, in turn, means that there is much room for improvement in the current methods for visual coding for machines.  
\end{abstract}


\thispagestyle{firstpage}

\section{Introduction}
Sensors on autonomous vehicles can generate up to 40 Gigabits per second (Gbps) worth of data~\cite{Gotz_data_AV}, depending on the settings. Yet, three orders of magnitude less data is sufficient to detect objects around the vehicle as accurately as when using all the data~\cite{MMSP2024_autonomous}. This is an example of the power of \emph{coding for machines}, an emerging technology whose goal is to compress data efficiently without compromising the accuracy of its subsequent analysis. It represents a departure from traditional compression techniques, which aim for efficient compression with the goal of input data recovery. In coding for machines, recovery of input data is not necessary - what matters is the accuracy of its subsequent analysis. 

Early ideas about coding for machines can be traced back to the famous book by Shannon and Weaver~\cite{Shannon-Weaver-1949}, where the \emph{semantic} communication problem was defined as follows: \emph{``How precisely do the transmitted symbols convey the desired meaning?''} Although the formulation is somewhat vague, one can notice that in such a communication scenario, recovery of transmitted symbols is not needed. What is important is that they convey the desired ``meaning,'' which we can interpret as the accuracy of subsequent analysis. Over the years, concrete examples of coding for machines have appeared in the research literature. For example, in~\cite{Chaddha1996}, the authors develop a scheme for joint image compression and classification, which allows image classification from the compressed representation. In~\cite{Ortega2000}, the authors develop compression methods for recognition and content-based retrieval, while~\cite{MVS_ICIP2011} presents compression of features that enable mobile visual search.  

Interest in the topic has increased considerably in recent years, fueled by rapid advances in deep learning and artificial intelligence (AI). As visual AI models continue to be deployed at increasing scales, questions like how much data they need and how will these data be delivered to them become crucial. This has led to the initiation of several standardization efforts in the area, such as JPEG AI~\cite{Alshina2024_MMM}, MPEG Video Coding for Machines (MPEG VCM)~\cite{VCM_overview2021}, and MPEG Feature Coding for Machines (MPEG FCM),\footnote{\url{https://www.mpeg.org/standards/MPEG-AI/4/}} 
both part of MPEG-AI.

In terms of research, coding for machines has been applied to various computer vision tasks, including image classification~\cite{Duan2022_PCS,Zhang2022_PCS,Nilesh_CVPR2023}, point cloud classification~\cite{Ulhaq2023,SPCGC_ICRA2024,Ulhaq2024}, object detection~\cite{Yuan2022_MIPR,Choi2022_TIP}, segmentation~\cite{Nilesh_CVPR2023,Choi2022_TIP,Fisher2025_TCSVT}, and so on. Coding for multiple tasks has also been considered~\cite{Alvar_ICIP2019,Matsubara_WACV2022}, particularly coding for human and machine vision~\cite{Choi2022_TIP,Yang2021_TMM,Ulhaq2024,SSSIC_TIP2021}. Coding for machines also plays an important role in (edge-cloud) collaborative intelligence~\cite{Kang2017}, split computing~\cite{Datta2022_ICPR}, Internet-of-Things (IoT)~\cite{Shlezinger2022}, and semantic communications~\cite{Getu2024_PIEEE}. In these related areas, the main research focus may be on other problems, such as dealing with noise in the communication channel or scheduling computation across heterogeneous devices, but compression of data for the purpose of subsequent analysis still plays an important role.  

In this paper we ask the following question: what is the best performance that \emph{any} data compression scheme can have on a given analysis task? Interestingly, despite the generality of the question, a precise answer can be given in many cases by exploiting the parallels between the outcome of the analysis and a discrete memoryless source (DMS) model from information theory~\cite{Cover_Thomas_2006}. Using this analogy, we provide Rate-Accuracy (RA) bounds for several popular problems in visual coding for machines, such as image classification, point cloud classification, and object detection. 

The paper is organized as follows. In Section~\ref{sec:DMS}, we review Rate-Distortion (RD) results for a DMS. In Section~\ref{sec:RA_bounds}, we explain the connection between DMS and coding for machines, and develop several RA bounds. In Section~\ref{sec:bounds-SOTA}, we examine how far are current approaches from the RA bounds, and in Section~\ref{sec:information_bottleneck}, we explore possible ways of getting closer to the bounds. Section~\ref{sec:conclusions} concludes the paper.

\section{Discrete Memoryless Source}
\label{sec:DMS}

A Discrete Memoryless Source (DMS) is a source of information that outputs symbols from a discrete set, say $\{1, 2, ..., K\}$, independently according to some probability distribution across the symbols. Let $T$ be the random variable that represents the symbols, $p(T)$ be the distribution over the source symbols, and let $t$ represent a specific symbol. The symbols are encoded by an encoder and then decoded by the decoder. Let $\widehat{T}$ be the random variable representing the output of the decoder, which may be different from the encoded symbol. Hamming distortion~\cite{Cover_Thomas_2006} is an indicator of the discrepancy between encoded and decoded symbols, that is, an error indicator: 
\begin{equation}
    d_H(t,\hat{t})=
    \begin{cases}
        0, &t=\hat{t},\\
        1, &t\neq \hat{t}.
    \end{cases}
    \label{eq:Hamming_distortion}
\end{equation}
The expected Hamming distortion is the expected fraction of decoded symbols that do not match the encoded symbols:  $D_H=\mathbb{E}\left[d_H(T,\widehat{T})\right]$.

The process of (potentially lossy) encoding and decoding of the symbols is represented as a conditional distribution $p(\widehat{T} | T)$. For a given distortion level $D$, the set of all feasible conditional distributions $p(\widehat{T}|T)$ is defined as:
\begin{equation}
    \mathcal{P}(D)=\left\{p(\widehat{T}|T)~:~\mathbb{E}\left[d_H(T,\widehat{T})\right]\leq D\right\}.
    \label{eq:quant_T}
\end{equation}
The Rate-Distortion (RD) function is defined as the minimum mutual information $I(T;\widehat{T})$ between encoder's input $T$ and decoder's output $\widehat{T}$ over the feasible set $\mathcal{P}(D)$:
\begin{equation}
    R(D) = \min_{p(\widehat{T}|T)~\in~\mathcal{P}(D)} I(T;\widehat{T}).
    \label{eq:RD_function_T}
\end{equation}
This represents a fundamental bound on lossy encoding of the outputs of the DMS: there are no encoding schemes that use a rate lower than $R(D)$ for encoding the outputs of a DMS while achieving distortion of at most $D$. Meanwhile, encoding at the rate $R(D)$ is asymptotically achievable when jointly encoding many outputs from a DMS~\cite{Cover_Thomas_2006}. 

It turns out that the closed form of the RD function for a DMS with uniform distribution $p(T)$ and Hamming distortion~(\ref{eq:Hamming_distortion}) is known. Specifically, for a DMS with $K$ equally likely symbols, the RD function is given by~\cite{mceliece1977}:
\begin{equation}
    R(D)=
    \begin{cases}
       \log_2(K) - D\log_2(K-1) - \mathcal{H}(D),\\  \hspace{90pt} \text{if} \enspace 0\leq D \leq 1-1/K,\\
       0, \hspace{81pt} \text{if} \enspace D \geq 1-1/K,
    \end{cases}
    \label{eq:RD_DMS}
\end{equation}
where $\mathcal{H}(D) = -D\log_2(D) - (1-D)\log_2(1-D)$ is the binary entropy function~\cite{Cover_Thomas_2006}. 

\section{Rate-Accuracy Bounds for Discrete Tasks}
\label{sec:RA_bounds}

Many vision tasks -- such as image or point cloud classification, segmentation, etc. -- have a discrete set of valid outputs, hence they can be considered classification problems. In this section, we show how Rate-Accuracy (RA) bounds can be established for these problems by exploiting the equivalence between rate-constrained classification and lossy encoding of a DMS, which was reviewerd in the previous section. 

First, note that we can interpret ``symbols'' of a DMS as \emph{class labels}, which are also discrete. Then $d_H$ in~(\ref{eq:Hamming_distortion}) is simply an indicator of a classification error, $D_H=\mathbb{E}\left[d_H(T,\widehat{T})\right]$ is the expected classification error, and $A=1-D_H$ is the (expected) classification accuracy.

\begin{figure}[t]
    \centering
    \includegraphics[width=\columnwidth]{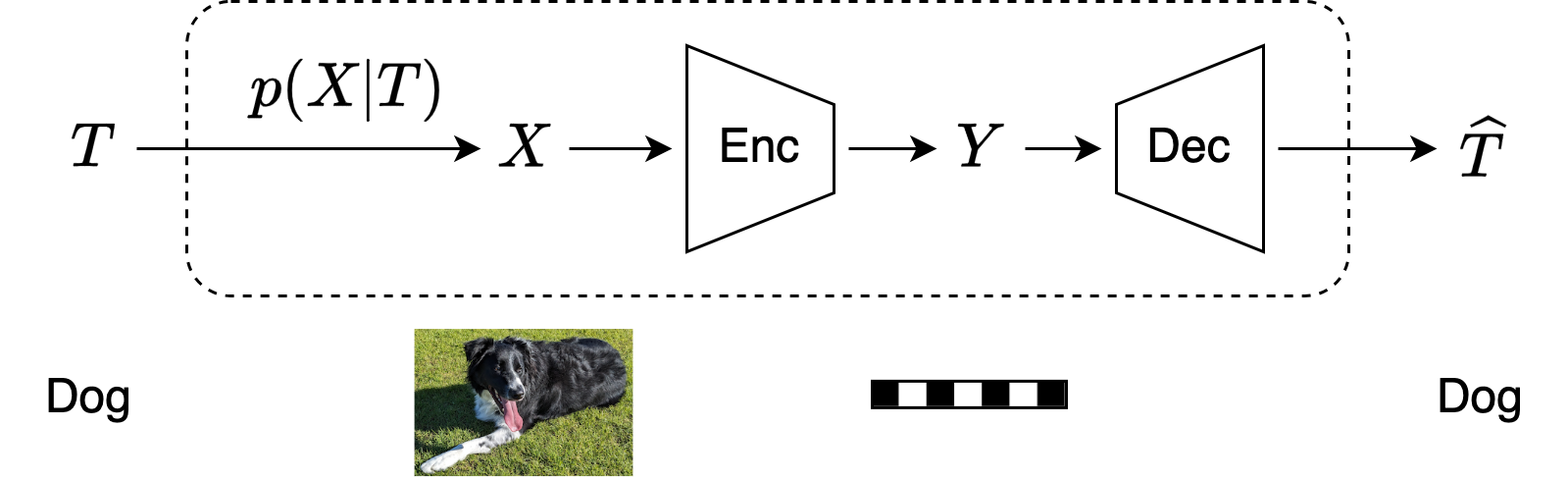}
    \caption{Coding for image classification as lossy encoding of a discrete memoryless source.} 
    \label{fig:classification_DMS}
\end{figure}

Conceptually, we can think of the processing chain 
$T \to X \to Y \to \widehat{T}$ shown in Fig.~\ref{fig:classification_DMS}, where $X$ is an image generated based on the class $T$ via $p(X|T)$ and $Y$ is the latent representation encoded at rate $R$. The overall mapping $T \to ... \to \widehat{T}$ represents lossy encoding of a DMS at rate $R$, hence we can use rate-distortion theory for a DMS to develop RA bounds for this type of system. More specifically, it should be clear that an image classifier with $K$ classes, where the image is encoded at $R$ bits on average, cannot be better than the rate-distortion bound for a $K$-symbol DMS at the same rate; otherwise, it would violate the well-established results on DMS. On the other hand, a perfect image classifier followed by an optimal $K$-symbol codec for a DMS could achieve the rate-distortion bound. Hence, the RD function for a DMS~(\ref{eq:RD_function_T}) essentially represents the RA bound for rate-constrained classification, with classification accuracy ($A$) and distortion ($D_H$) related by $A=1-D_H$. 

Note that $0\leq D_H\leq1$ and $0 \leq \mathcal{H}(D_H) \leq 1$. Hence, when the number of classes is large ($K \gg 1$),  $\mathcal{H}(D_H)$ becomes insignificant compared to the other terms, and the following approximations become valid for $0\leq D_H < 1-1/K$:
\begin{equation}
\begin{split}
    R(D_H) &\approx \log_2(K) - D_H\log_2(K-1)\\
    &\stackrel{\text{(a)}}{\approx} (1-D_H) \log_2(K)\\
    &\stackrel{\text{(b)}}{=}A \log_2(K),
\end{split}
\end{equation}
where (a) follows from $K \gg 1$ and (b) follows from $A=1-D_H$. Hence, the relationship between the accuracy and rate becomes approximately linear, and the derivative of the accuracy with respect to rate is approximately
\begin{equation}
    A'(R) \approx \frac{1}{\log_2(K)}.
    \label{eq:accuracy_derivative}
\end{equation}
This represents the expected improvement in accuracy for each additional bit. It is a rate-accuracy analogue of the classic rate-distortion result (``rule of thumb''), which says that the mean squared error distortion for a Gaussian source reduces by a factor of 4 (or 6~dB) for each additional bit~\cite{Cover_Thomas_2006}. Here, in the limit, classification accuracy should improve by $1/\log_2(K)$ for each additional bit.

\begin{figure}
    \centering
    \includegraphics[width=\columnwidth]{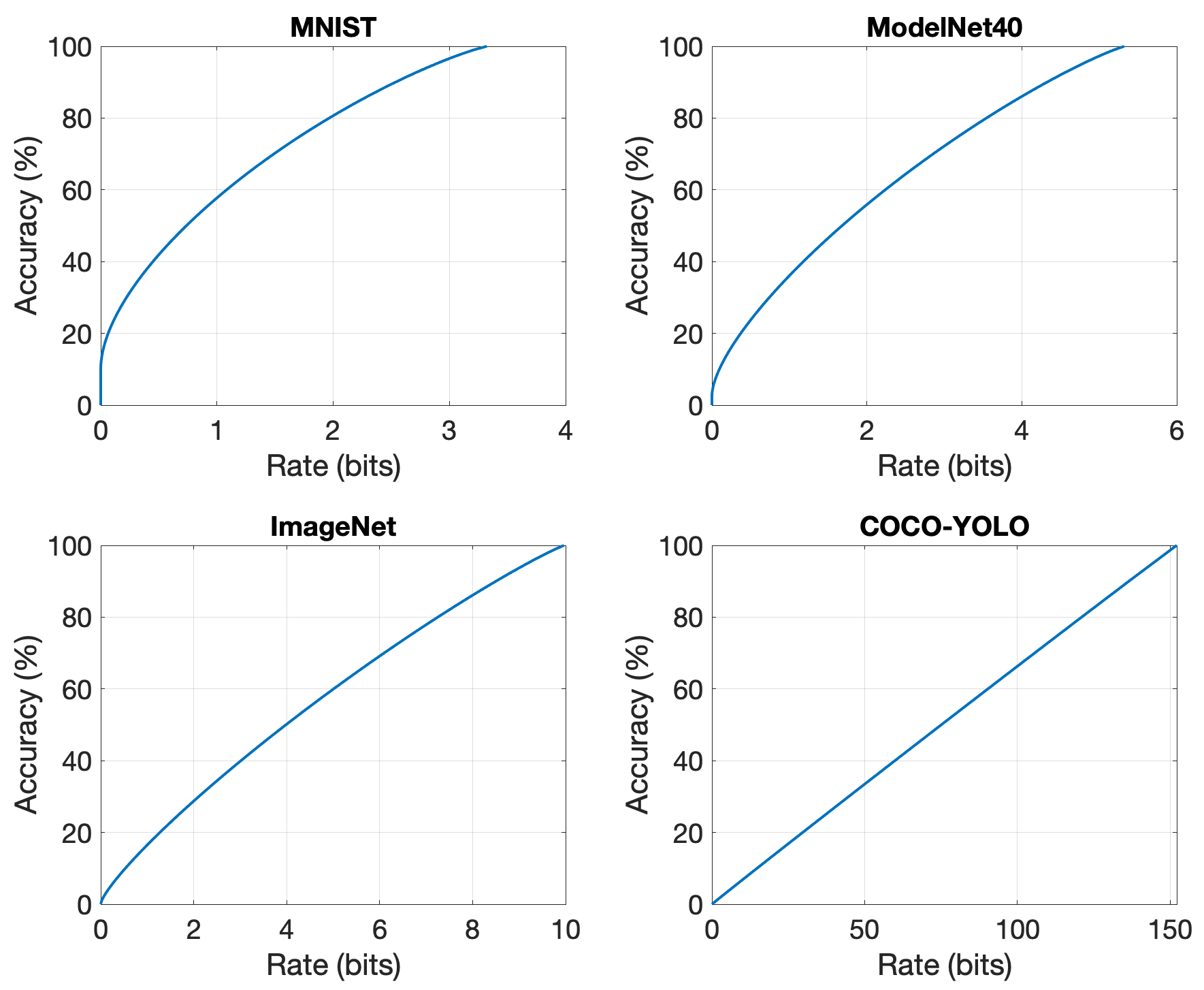}
    \caption{Rate-accuracy bounds for popular datasets, assuming uniform distribution across classes}
    \label{fig:AR_plots}
\end{figure}

Fig.~\ref{fig:AR_plots} shows the RA bounds for classification on MNIST ($K=10$), ModelNet40 ($K=40$) and ImageNet ($K=1000$), assuming equiprobable classes and using~(\ref{eq:RD_DMS}). The bottom-right plot in Fig.~\ref{fig:AR_plots} shows rate-accuracy plot for YOLO-like object detection~\cite{YOLO_CVPR2016} on the COCO dataset~\cite{COCO}. To create this plot, object detection was modeled as classification in the following way. It was assumed that there are between 1 and 15 objects in an image, that each object can be in one of 98 positions\footnote{YOLO considers 98 anchor positions per image for object bounding boxes.} and that each object can be any of the 80 COCO classes. Then the number of possible object configurations in an image is
\begin{equation}
    K=\sum_{i=1}^{15} \binom{98}{i}80^i \approx 6.4\cdot10^{45}.
\end{equation}
If each object configuration is considered as a separate class, then object detection can be modeled as a classification problem. The plot in Fig.~\ref{fig:AR_plots} assumes that all classes are equiprobable. Note that the accuracy of object detection is usually quantified in terms of Intersection-over-Union (IoU) or mean Average Precision (mAP) rather than classification accuracy, and the plot of such metrics against bitrate may be different.

\begin{figure}
    \centering
    \includegraphics[width=0.7\columnwidth]{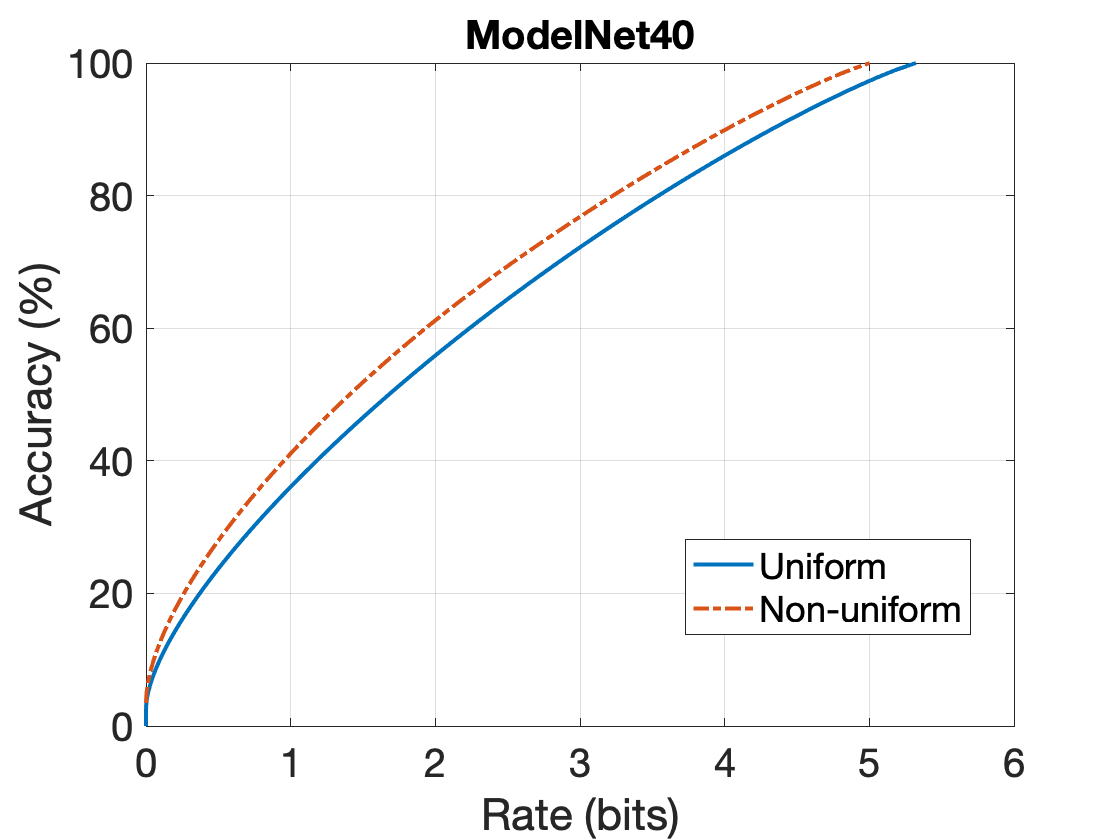}
    \caption{Rate-accuracy bounds for ModelNet40 with uniform and non-uniform source distribution}
    \label{fig:ModelNet40}
\end{figure}

What if classes are not equiprobable? In that case, there is no simple closed-form equation for the RD function like~(\ref{eq:RD_DMS}), but we can use the Blahut-Arimoto algorithm~\cite{Cover_Thomas_2006} to compute it. Fig.~\ref{fig:ModelNet40} shows two curves for ModelNet40: one is the RA bound assuming a uniform distribution $p(T)$ across the classes, which is given by~(\ref{eq:RD_DMS}), and the other is the RA bound calculated using the Blahut-Arimoto algorithm on the non-uniform distribution $p(T)$ that follows the ModelNet40 test set. As seen in the figure, the bound is better for the non-uniform distribution, meaning that fewer bits are needed to achieve the same accuracy, compared to a source with uniform distribution. It is well known that uniform distribution has the highest entropy among all distributions over the same sample space~\cite{Cover_Thomas_2006}. Fig.~\ref{fig:ModelNet40} illustrates the same principle in the lossy case, where the uniform distribution has the worst RA bound among all distributions over the same sample space.

\section{How far are we from the bounds?}
\label{sec:bounds-SOTA}

Now that we know the RA bounds for several popular problems in coding for machines, we can ask how far are the current state-of-the-art (SOTA) results from these RA bounds? 

\textbf{MNIST image recognition:} This is a relatively old problem,\footnote{\url{https://en.wikipedia.org/wiki/MNIST_database}} which has been essentially solved even prior to the modern era of deep learning. For example, the Support Vector Machine (SVM) method from~\cite{Decoste_ML_2002} achieves an error rate of 0.56\% (accuracy of 99.44\%) with computational requirements far below modern neural networks. In the more recent literature on coding for machines, the authors in~\cite{lossy_comp_neurips2021} have reported their best result obtained by a Variational Invariant Compressor (VIC) as 99.1\% accuracy at 5.7 bits/image.  

On the other hand, using the reasoning presented in the previous section, we could create a low-complexity encoder as follows: classify the image using the SVM from~\cite{Decoste_ML_2002} to obtain the digit (class), then encode the digit using a fixed-length code. Since there are 10 possible digits at the output of the SVM, 4 bits would suffice to encode the output of the SVM, since $2^4=16>10$. Hence, we can easily obtain the accuracy of 99.44\% at 4 bits/image. Let us call this \underline{Method~1}. 

On the other hand, we could try to create a variable-length code for encoding the outputs of the classifier (the digits). Even though the digits are assumed to be equally likely -- which would suggest a fixed-length code is optimal -- the fact that the length of each codeword has to be an integer forces us to use 4-bit codewords and therefore end up with 16 possible codewords (in the fixed-length code), even though the number of digits is only 10. If we run Huffman code design for this case, we end up with a variable-length code shown in Table~\ref{tab:Huffman}, which has the average codeword length of 3.6 bits. Hence, in this case we have the accuracy of 99.44\% at 3.6 bits/image. Let us call this \underline{Method~1a}.   

Finally, if we encode three images at a time, we could do even better. There are $10^3$ triplets of digits that we could obtain by processing three input images by the SVM, so we could use a fixed-length code with 10 bits to encode the triplet, since $2^{10} = 1024 > 1000 = 10^3$. Hence, we would be using $10/3 \approx 3.33$ bits/image and still having 99.44\% accuracy. Let us call this \underline{Method~2}. Note that this method is fairly close to the entropy of the source, which is $\log_2{10}\approx3.32$ bits/digit.

\begin{figure}[t]
    \centering
    \includegraphics[width=0.7\columnwidth]{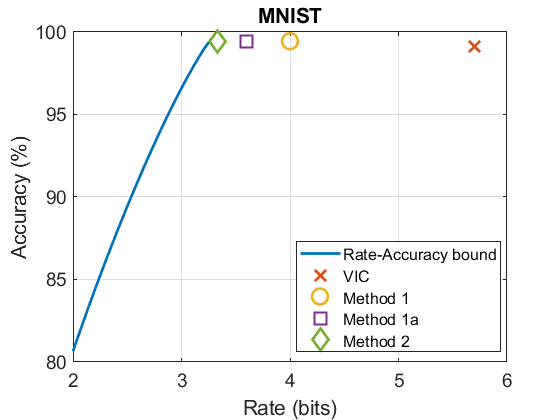}
    \caption{Distance of VIC method from~\cite{lossy_comp_neurips2021} and the three methods described in this paper from the RA bound for MNIST. Rate is in bits per image.}
    \label{fig:MNIST-distance}
\end{figure}

The performance of these three methods, as well as that of VIC~\cite{lossy_comp_neurips2021}, is shown in Fig.~\ref{fig:MNIST-distance}. All four methods perform relatively close to the RA bound, with Method 2 coming especially close. Part of the reason for such success is that MNIST recognition is a ``solved'' problem with few classes, all equally likely, hence a lightweight classifier followed by fixed-length coding is already close to optimal. The situation becomes much more challenging for more complicated tasks, as we will see next. 

\begin{table}[t]
    \centering
    \caption{Huffman code for the digits}
    \begin{tabular}{c|l}
        \toprule
        Digit & Codeword \\
        \midrule
         0 & 00 \\
         1 & 01 \\
         2 & 1000 \\
         3 & 1001 \\
         4 & 1010 \\
         5 & 1011 \\
         6 & 1100 \\
         7 & 1101 \\
         8 & 1110 \\
         9 & 1111 \\
         \bottomrule
    \end{tabular}
    \label{tab:Huffman}
\end{table}

\begin{figure}[t]
    \centering
    \includegraphics[width=0.7\columnwidth]{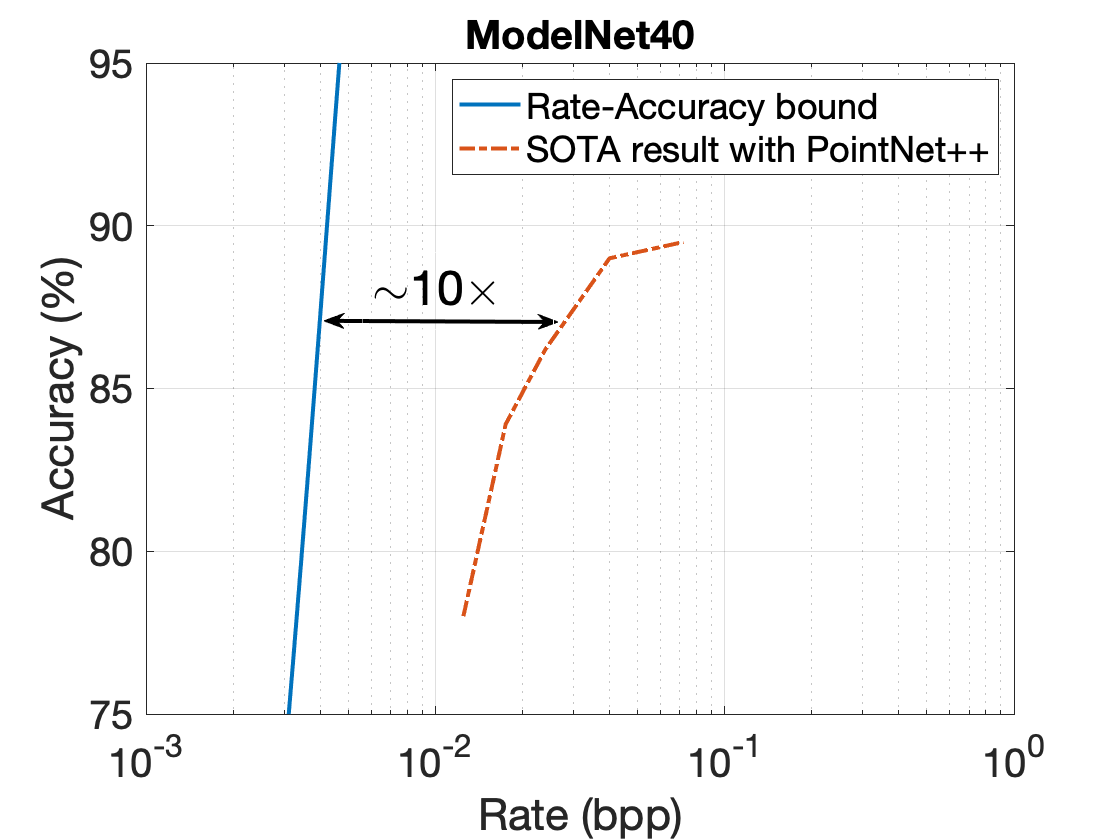}
    \caption{Distance of a SOTA method from the RA bound for ModelNet40. Rate in bits per point (bpp) is computed for the point cloud size of 1024. }
    \label{fig:ModelNet40-distance}
\end{figure}

\textbf{ModelNet40 point cloud classification:} Although SOTA results are continuously improving, at the time of writing this article, it appears that the best results on the rate-constrained point cloud classification on ModelNet40 are those in~\cite{Ulhaq2024}. Fig.~\ref{fig:ModelNet40-distance} shows the RA bound on this problem along with the best result from~\cite{Ulhaq2024}. In this plot, rate is shown in terms of bits per point (bpp), which is calculated assuming that each point cloud has 1024 points. As seen in the figure, the distance between the two curves varies depending on where exactly it is measured, but at the higher end of the curve (higher accuracies), it is at least one order of magnitude (10$\times$). Hence, the current SOTA scheme for point cloud compression for classification uses at least an order of magnitude more bits than what is theoretically the best possible. 

\begin{figure}[t]
    \centering
    \includegraphics[width=0.7\columnwidth]{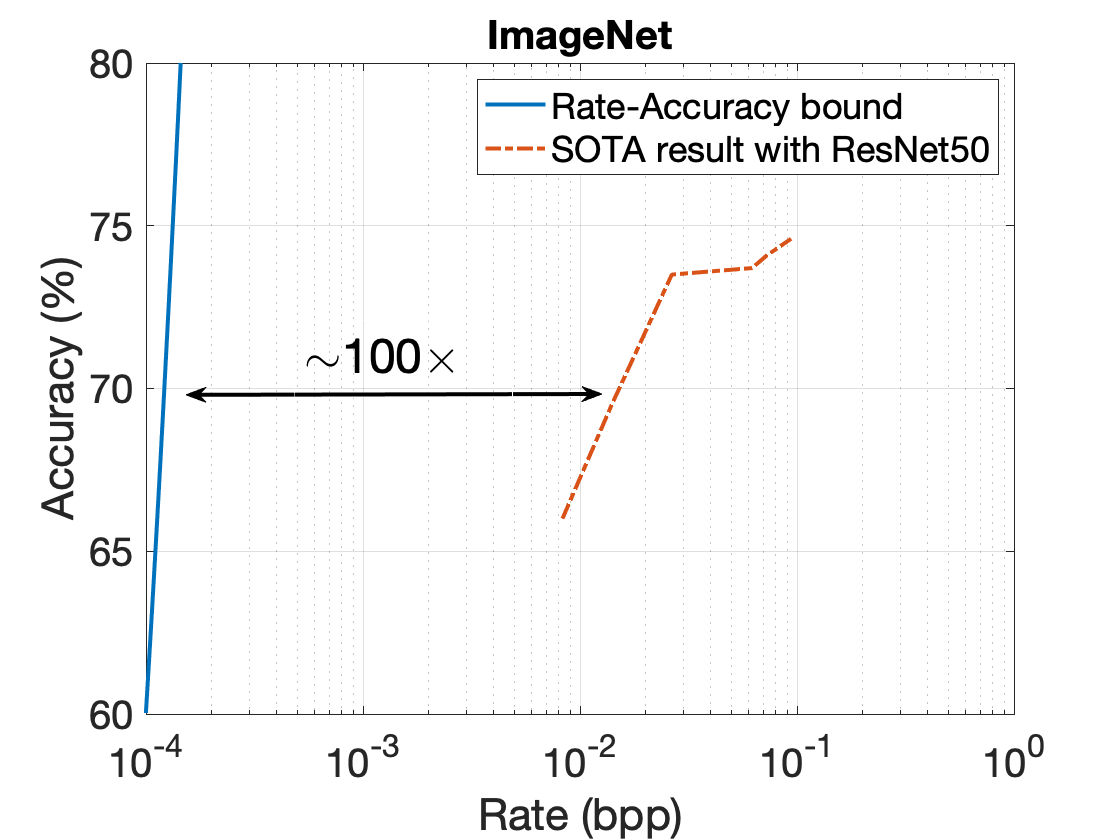}
    \caption{Distance of a SOTA method from the RA bound for ImageNet. Rate in bits per pixel (bpp) is computed for the image resolution of 224$\times$224.}
    \label{fig:ImageNet-distance}
\end{figure}

\textbf{ImageNet image classification:} Next, we examine rate-constrained image classification using ResNet50~\cite{ResNet_CVPR2016} on ImageNet~\cite{ImageNet}, where the current SOTA results come from~\cite{Harell2025}. The RA bound and the SOTA result are shown in Fig.~\ref{fig:ImageNet-distance}. Here, again, the actual distance between the curves depends on where it is measured, but a the higher end of the curve, the distance is at least two orders of magnitude (100$\times$).  

\textbf{COCO object detection:} Finally, we look at object detection. The RA bound on YOLO-like object detection on the COCO dataset~\cite{COCO} in Fig.~\ref{fig:AR_plots} was derived by considering object detection as a classification problem over the classes defined by the various configurations of object locations and object classes in an image. However, the established way of measuring the accuracy of object detection models in the literature is via mean Average Precision (mAP), which considers the overlap of detected and ground-truth object boxes, as well as their classes. Since there is no obvious way to convert between mAP and classification accuracy as defined above, we cannot show a plot as we did for ModelNet40 and ImageNet. But we can get a rough estimate of how far SOTA methods are from the RA bound as follows. According to Fig.~\ref{fig:AR_plots}, perfect accuracy of object detection in the COCO-YOLO framework should be achievable with 150 bits per image. Assuming image resolution of 640$\times$480, which is a common resolution in the COCO dataset, this is approximately $5\cdot10^{-4}$ bits per pixel (bpp). Meanwhile, the SOTA result~\cite{Harell2025} on the COCO dataset using YOLOv3~\cite{YOLOv3} achieves mAP of about 55.5\% at $0.5$ bpp. Hence, even at three orders of magnitude higher bitrate, the accuracy is still far from perfect. This suggests that the current SOTA methods are at least three orders of magnitude (1000$\times$) away from RA bounds in object detection.

\section{How to approach the RA bounds?}
\label{sec:information_bottleneck}

\subsection{Removal of irrelevant information}
\begin{figure}[t]
    \centering
    \includegraphics[width=0.8\columnwidth]{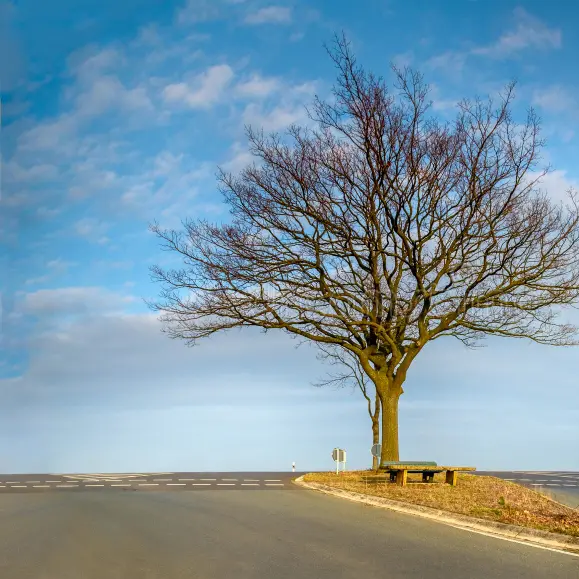} 
    \includegraphics[width=0.8\columnwidth]{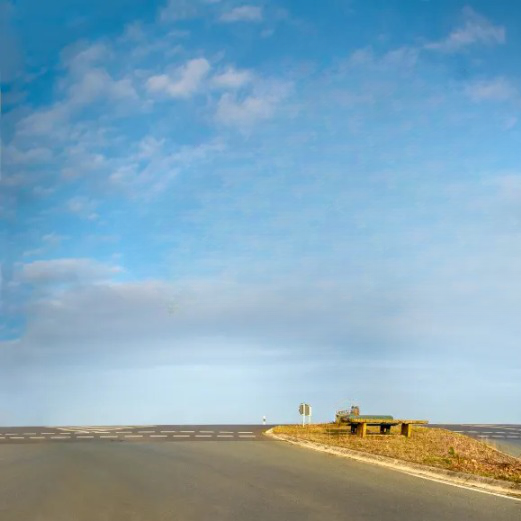}
    \caption{Top: An image of a roadway. Bottom: The same image, but with some of the information that is irrelevant to counting vehicles removed.}
    \label{fig:irrelevant_info}
\end{figure}

Conventional codecs try to minimize the rate at which the input $X$ is encoded,  with the aim of recovering a good approximation $\widehat{X}$ to the input $X$. They do this by removing the statistical redundancy from the source $X$. Codecs for machines also need to remove statistical redundancy. But, in order to be efficient, they also need to remove information that is irrelevant to the task $T$. An example of \emph{irrelevant information} is shown in Fig.~\ref{fig:irrelevant_info}. Suppose the task is to count the vehicles passing through the road segment shown in the image. For that task, the tree shown in the top image is irrelevant.\footnote{The tree is not redundant, because it is the only tree in the image, but it is irrelevant to the task of counting vehicles.} In the bottom image, that tree has been removed. It should be intuitively clear that encoding the bottom image would require fewer bits than encoding the top one, because the bottom image no longer has the fine textures of the tree branches. 

Removal of irrelevant information is the key difference between codecs for machines and conventional codecs. Indeed, it is also the main reason why codecs for machines are able to deliver the same task accuracy at much lower bitrates than conventional codecs. The question is - given a task, how to remove information that is irrelevant to the task? As will be seen next, neural networks can provide a way to do that.

Consider a feedforward network trained to accomplish a certain task, for example image classification. Let the processing chain of that network be represented by $X \to Y^{(1)} \to Y^{(2)} \to ... \to Y^{(L)} \to T$, where $X$ is the input (in our example, an image), $L^{(i)}$ is the feature tensor at the output of the $i$-th layer, and $T$ is the task output (in our example, class label). In a feedforward network, each layer's output is computed from the previous layer's output, so $Y^{(i)} \to Y^{(j)} \to Y^{(k)}$ forms a Markov chain for any $i<j<k$. Based on the Data Processing Inequality (DPI)~\cite{Cover_Thomas_2006}, we have that $I(Y^{(i)};Y^{(j)}) \geq I(Y^{(i)};Y^{(k)})$, with equality if and only if the mappings between the layers are invertible. Hence, information about earlier layers (as well as the input) is removed by non-invertible layer processing.  

What about skip connections, which are also common in feedforward networks? Consider a processing chain shown below:
\begin{singlespace}
\begin{equation}
\begin{tikzpicture}[
  inner sep=0pt,
  outer sep=0pt,
  baseline=(x.base),
]
  \tikzstyle{arrow} = [thin,->,>=stealth]
  
  \path[every node/.append style={anchor=base west}]
    (0, 0)
    \foreach \name/\code in {
      z/ Y^{(i)}~,
      tmp/\,\qquad,
      x/ ~Y^{(j)}~,
      tmp/\,\qquad,
      t/ ~Y^{(k)}~,
      tmp/\,\qquad,
      y/ ~Y^{(l)}%
    } {
      node (\name) {$\code$}
      (\name.base east)
    }
  ;
  \path[
    every node/.append style={
      anchor=base,
      font=\scriptsize,
    },
  ]
    (x.base) -- node[above=2.0\baselineskip] (f) {} (y)
    (x.base) -- node[below=0.2\baselineskip] (g) {} (t)
    (t.base) -- node[below=0.2\baselineskip] (h) {} (y)
  ;
  \draw [arrow] (z) -- (x);
  \draw [arrow] (x) -- (t);
  \draw [arrow] (t) -- (y);
    
  \begin{scope}[
    >={Stealth[length=5pt]},
    thin,
    rounded corners=2pt,
    shorten <=.3em,
    shorten >=.3em,
  ]
    
    \def\GebArrow#1#2#3{
      \draw[->]
        (#2.south) ++(0, -.3em) coordinate (tmp)
        (#1) |- (tmp) -| (#3)
      ;%
    }
    \GebArrow{x}{f}{y}
  \end{scope}
\end{tikzpicture}
\label{eq:skip_connection}
\end{equation}
\end{singlespace}
\noindent where $i<j<k<l$ and $Y^{(l)}$ is a function of $Y^{(j)}$ and $Y^{(k)}$, which could include concatenation, addition, etc. In this case, $Y^{(j)} \to Y^{(k)} \to Y^{(l)}$ is not a Markov chain because, given $Y^{(k)}$, $Y^{(l)}$ is not conditionally independent of $Y^{(j)}$. However, $Y^{(i)} \to Y^{(j)} \to Y^{(l)}$ is a Markov chain, and the DPI shows that $I(Y^{(i)};Y^{(j)}) \geq I(Y^{(i)};Y^{(l)})$. Hence, DPI acts ``across'' skip connections, but not necessarily ``under'' them. 

The arguments presented above show that feedforward networks provide many opportunities for information to be removed from the input. But how do we remove \emph{irrelevant} information? We do this by training the network to be sufficiently accurate\footnote{The exact meaning of ``sufficiently accurate'' depends on the application.} in performing the task. If the network is sufficiently accurate, then task-relevant information must be making its way to the network's output. This, in turn, means that information that was removed via processing by various non-invertible layers must be irrelevant to the task! 

\begin{figure}
    \centering
    \includegraphics[width=0.5\linewidth]{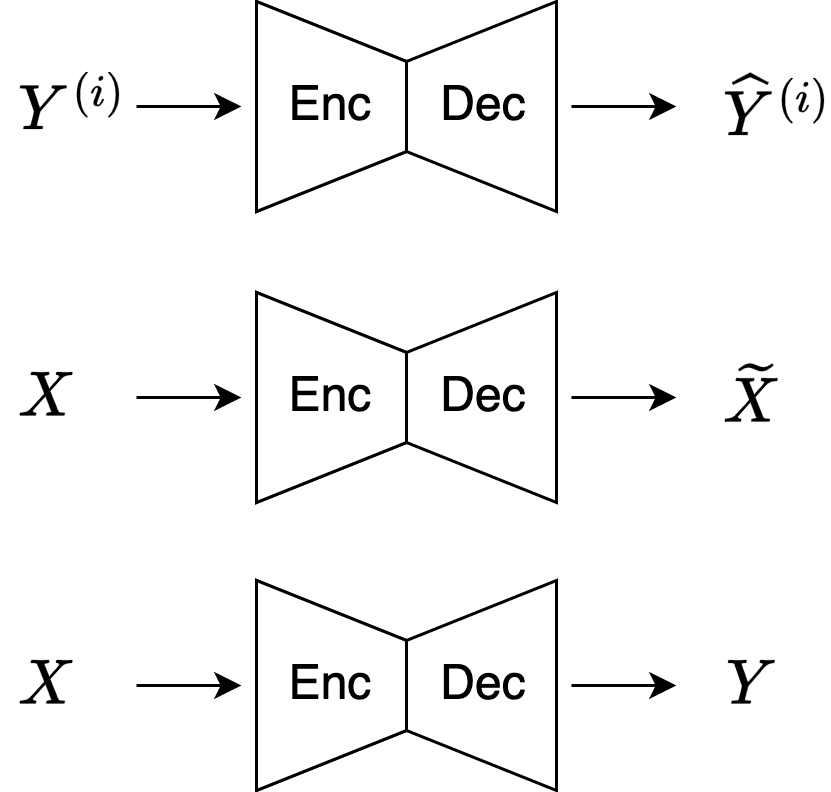}
    \caption{Three types of codecs for machines}
    \label{fig:3_codecs}
\end{figure}

This line of reasoning suggests that a good codec for machines may be constructed by compressing the features produced by a certain layer $Y^{(i)}$ in a neural network trained for a particular task, as shown at the top of Fig.~\ref{fig:3_codecs}. In this design, the role of the initial layers of the network ($X \to Y^{(1)} \to Y^{(2)} \to ... \to Y^{(i-1)}$) is to remove some of the irrelevant information, while the goal of the codec is to remove statistical redundancy. The decoded features ($\widehat{Y}^{(i)}$) are then fed to the remainder of the network for analysis. This strategy is being pursued in MPEG Feature Coding for Machines (MPEG FCM).  

Another strategy is to modify a conventional codec, say by pre-processing or by adjusting bit allocation, in order to remove some of the irrelevant information in addition to statistical redundancy. For example, such a codec could input an image ($X$) and decode an image ($\widetilde{X}$) with some of the irrelevant information removed or suppressed, as shown in the middle of Fig.~\ref{fig:3_codecs}. Then $\widetilde{X}$ would be subject to further analysis.  This type of strategy is being pursued in MPEG Video Coding for Machines (MPEG VCM)~\cite{VCM_overview2021}.

Finally, the third strategy is to develop a different kind of codec, which could, for example, input an image ($X$) but decode features ($Y$) for subsequent analysis, as shown in the bottom of Fig.~\ref{fig:3_codecs}. This type of codec tries to simultaneously remove irrelevant information and statistical redundancy. Examples of designs of such codecs include~\cite{Choi2021_ICIP}, where it was used for the base layer of a human-machine scalable coding system,  and~\cite{Matsubara_WACV2022}, where such codec was referred to as \emph{entropic student}. But the foundation of such coding has been known in the literature as \emph{information bottleneck}~\cite{IB_Allerton1999}, and will be briefly reviewed in the next section. 

Ultimate performance bounds for the three types of codecs shown in Fig.~\ref{fig:3_codecs} were studied in~\cite{Harell2025} using RD theory. The results show that all three types of codecs have the same RD bounds. That is to say, the best possible performance that can be achieved with each of the three designs shown in Fig.~\ref{fig:3_codecs} is the same, and neither design is inherently superior to the others. This means that the choice of the codec for a particular application should be made based on the actual performance under other design considerations: compute and energy requirements, hardware availability, cost, privacy, security, etc.

\subsection{Information bottleneck}
The notion of removal of irrelevant information is captured by the concept of information bottleneck (IB)~\cite{IB_Allerton1999}. Consider the processing chain $X\to Y \to T$. IB is defined as 
\begin{equation}
    \min_{p(Y | X)} \quad I(X; Y) - \beta \cdot I(Y; T),
\label{eq:IB}
\end{equation}
where $I(\cdot;\cdot)$ is the mutual information~\cite{Cover_Thomas_2006}, $p(Y | X)$ is the mapping from the input $X$ to the representation $Y$, and $\beta>0$ is the IB Lagrange multiplier~\cite{IB_Allerton1999}. IB seeks to find the transformation of input $X$ into representation $Y$ that removes all information from $X$ (minimizes $I(X; Y)$) except the information relevant to task $T$ (maximizes $I(Y; T)$). In other words, IB seeks to remove from $X$ information that is irrelevant to $T$. Note that removal of statistical redundancy is implicit in minimizing $I(X; Y)$. 

Although IB provides a good conceptual framework for codecs for machines, practical codec designs directly based on~(\ref{eq:IB}) are difficult, except in special cases. This is due to the challenges associated with estimating mutual information in high-dimensional spaces. In the case of visual codecs, input $X$, latent representation $Y$, and sometimes the task output $T$, are all high-dimensional. Instead, one simplification is to consider a deterministic mapping $X \overset{g_a}{\to} Y$, where $g_a$ is referred to as the \emph{analysis transform} and often includes quantization. With $H(\cdot)$ and $H(\cdot | \cdot)$ representing the entropy and conditional entropy, respectively, we now have $H(Y | X) = 0$ because, when $X$ is known, there is no uncertainty in $Y$. Hence, $I(X; Y) = H(Y) - H(Y | X) = H(Y)$. Estimating $H(Y)$ is still challenging, but less so than estimating $I(X; Y)$. Moreover, there has been tremendous progress in latent-space entropy estimation in recent years~\cite{he2022elic,jiang2023mlic,NeuralEntropy_CVPR2024,Fourier_PCS2024}. Hence, there are established methods for taking care of the first term in~(\ref{eq:IB}) when $H(Y | X) = 0$.

The second term in~(\ref{eq:IB}) is meant to ensure that the task accuracy remains high. Again, due to the challenges of estimating $I(Y; T)$, it is common to replace this term with another function that is simpler to compute and more directly related to the task accuracy. Several options have been presented in the literature. If the task labels are available (supervised case), it is common to optimize the mapping from $Y$ to $T$, denoted $g_t$, using the task loss $\mathcal{L}(g_t(Y),T)$ in place of $-I(Y; T)$. In this case, IB becomes 
\begin{equation}
    \min_{g_a,g_t} \quad H(Y) + \beta \cdot \mathcal{L}(g_t(Y), T).
\label{eq:IB_supervised}
\end{equation}

If the task labels are not available, or if a design goal is to utilize a pre-trained baseline model to perform the task, then the following approach can be used. Let $Y^{(i)}$ denote the features at the $i$-th layer of the baseline model, and let $g_s$ be the synthesis transform (latent space transform~\cite{Choi2021_ICIP}, sometimes called a \emph{bridge}~\cite{Seleem2024PC_JIVP}) whose goal is to approximate $Y^{(i)}$ from $Y$. Then the IB becomes
\begin{equation}
    \min_{g_a,g_s} \quad H(Y) + \beta \cdot d(g_s(Y), Y^{(i)}),
\label{eq:IB_unsupervised}
\end{equation}
where $d(g_s(Y), Y^{(i)})$ is some measure of distortion between $g_s(Y)$ and $Y^{(i)}$, for example, mean squared error (MSE), mean absolute error (MAE), Kullback-Leibler (KL) divergence, cosine similarity, etc. Since task labels are not involved in~(\ref{eq:IB_unsupervised}), this form of IB can be trained in an unsupervised manner and therefore on much larger datasets.

Results in~\cite{Harell2025} indicate that the best RD performance is achievable in the supervised case, when labels are available. Among unsupervised cases, matching (distilling) deeper layers generally leads to better RD performance~\cite{Harell2025}. All of these simplifications represent departures from the original IB~(\ref{eq:IB}) in favor of more practical systems that can be developed and trained using existing technology and datasets. The success of the systems presented in~\cite{Choi2021_ICIP,Matsubara_WACV2022,Choi2022_TIP}, which were built using IB principles, showcases the potential of this approach in getting closer to the ultimate rate-accuracy bounds.

\section{Conclusions}
\label{sec:conclusions}
In this paper we drew parallels between coding for machines and the Discrete Memoryless Source (DMS) model from information theory. This analogy allowed us to compute Rate-Accuracy (RA) bounds -- the minimum number of bits required for a certain level of accuracy -- for several popular problems in visual coding for machines. By comparing these RA bounds with the best current results in the literature on specific problems, it was revealed that the current state-of-the-art is at least an order of magnitude, and in some cases several orders of magnitude, away from the RA bounds. This means that there is much room for improvement in coding for machines. 

Possible strategies for approaching the bounds were discussed and related to the relevant coding standards in the field. In particular, the notion of removal of irrelevant information was discussed as a key component of coding for machines. Further, several high-level designs of codecs for machines were described, with particular emphasis on the designs inspired by the concept of information bottleneck.

\section{Acknowledgment}
The author is grateful to the former student Mateen Ulhaq for providing an implementation of the Blahut-Arimoto algorithm.

\bibliographystyle{IEEEbib}
\bibliography{refs}

\end{document}